# Effects of Layer Stacking on the Combination Raman Modes in Graphene

*Rahul Rao,[1][*] Ramakrishna Podila,[2] Ryuichi Tsuchikawa,[3] Jyoti Katoch,[3] Derek Tishler,[3] Apparao M. Rao,[2] Masa Ishigami[3]*

[1]Materials and Manufacturing Directorate, Air Force Research Laboratory, WPAFB, 45433

[2]Department of Physics and Astronomy, Clemson University, Clemson, SC, 29634

[3]Department of Physics and Nanoscience Technology Center, University of Central Florida, Orlando, FL, 32816

Email: rahul.rao@wpafb.af.mil

**Abstract**

We have observed *new* combination modes in the range from 1650 – 2300 cm$^{-1}$ in single- (SLG), bi-, few-layer and incommensurate bilayer graphene (IBLG) on silicon dioxide substrates. The M band at ~1750 cm$^{-1}$ is suppressed for both SLG and IBLG. A peak at ~1860 cm$^{-1}$ (iTALO$^{-}$) is observed due to a combination of the iTA and LO phonons. The intensity of this peak decreases with increasing number of layers and this peak is absent in bulk graphite. Two previously unidentified modes at ~1880 cm$^{-1}$ (iTALO$^+$) and ~2220 cm$^{-1}$ (iTOTA) in SLG are tentatively assigned as combination modes around the K point of the graphene Brillouin zone. The peak frequencies of the iTALO$^+$ (iTOTA) modes are observed to increase (decrease) linearly with increasing graphene layers.

---
[*] To whom correspondence should be addressed

Single layer (SLG) and bi-layer graphene (BLG) have recently attracted much attention from the research community, mainly due to their extraordinary electronic properties, which are interesting for both fundamental and applied sciences.[1, 2] SLG and BLG are quite different from each other with respect to their band structure. SLG is a semimetal with a vanishing density of states at the Fermi level,[3] while AB-stacked BLG possesses massive Dirac fermions with a transverse field-tunable band gap.[4] On the other hand, incommensurate BLG (IBLG) behaves in a similar fashion as SLG with reduced Fermi velocities.[5, 6]

Raman spectroscopy is the standard technique to distinguish between SLG, BLG, IBLG, and graphene samples with a few layers (FLG).[7] The most commonly used Raman signature for layer thickness is a peak occurring at ~2700 cm$^{-1}$ called the G' (also called the 2D) band, which is an overtone of the disorder-induced D band located at ~1350 cm$^{-1}$. Both the D and G' bands occur due to an inter-valley double resonance Raman process[8] where the D band phonon scattering is a second order process mediated by a defect, while the G' band occurs due to scattering by two phonons and does not need any defects for activation. The G' band in SLG can be fit to a single Lorentzian peak and its intensity has been found to be much higher than that of the G band (located at ~1580 cm$^{-1}$) for SLG; hence it is often used as an indicator of an SLG region.[7, 9-12] On the other hand, the G' band in BLG can be clearly deconvoluted into four Lorentzian peaks and its intensity is lower than that of the G' band on silicon dioxide.[7] As the number of layers increase to more than 3 the G' band evolves into a two-peak structure along with a concomitant decrease in intensity with respect to the G band. It has recently been shown that IBLG can be distinguished between SLG and BLG by the presence of new defect-induced peak (I band) located on the high frequency side of the D band (Fig. 1a).[13] The I band appears due to one layer imposing a perturbation on the other and is a signature for the presence of non-AB stacked graphene. Moreover, the frequency of the I band depends on the angle of orientation between the folded and parent graphene layer.[13] In another recent report, other weak intensity peaks between 1650 and 2150 cm$^{-1}$ have been observed from SLG and IBLG.[14] These weak intensity peaks were assigned to combination modes that occur due to a double resonance Raman scattering process involving the iTO, LO and iTA phonons.[14]

We have performed detailed investigations of the combination modes involving iTO, LO and TA phonons in SLG, BLG, FLG and IBLG and report three new features in the region between 1650 and 2300 cm$^{-1}$. (1) We observe a previously unidentified dispersive mode at ~1880 cm$^{-1}$ (iTALO$^+$) when excited with $E_{laser}$ = 2.33 eV in SLG, which strongly depends on the number and stacking order of graphene layers. This mode is tentatively assigned as a combination of the oTO + LO phonons mode around the K point in the graphene Brillouin zone. (2) Another previously unidentified mode is observed at ~2220

cm$^{-1}$ in SLG (when excited with $E_{laser}$ = 2.33 eV). This mode has a negative dispersion with respect to laser energy and is tentatively assigned as a combination of the iTO and iTA phonons (iTOTA mode) around the K point. (3) The combination modes involving the LO phonon (iTALO$^-$, iTALO$^+$, and LOLA modes) upshift in frequency with increase in the number of graphene layers, while the iTOTA mode frequency downshifts with increasing graphene layers. An additional stiffening of all the combination modes is observed for IBLG.

**Results and Discussion**

Figure 1a shows the D and G band region and the G' band regions from the graphene samples collected using $E_{laser}$ = 2.33 eV. Also included in Fig. 1 are spectra collected from bulk graphite (HOPG). The D band intensity is very low across all graphene samples and is negligible for HOPG. Not surprisingly, the $I_D/I_G$ value for IBLG is the highest and it decreases in general as the number of layers increase as shown in the inset in Fig. 1a. In addition, a second peak in the D band region can be observed in the IBLG spectrum. This peak, called the I band, appears at 1374 cm$^{-1}$ and can be used as a metric for identification of IBLG (see Supporting Information Fig. S1 for a magnified view of the I band). The G' band from SLG, BLG, FLG and HOPG can be fit to 1, 4, 2 and 2 Lorentzian peaks, respectively, thus confirming the presence of 1, 2, few layer graphene and bulk graphite (Fig. 1b). The Raman signature from IBLG is different from both SLG and BLG, where the G' band intensity is higher than the G band, but reverts to a single Lorentzian peak similar to SLG with a blue-shifted (~ 7 cm$^{-1}$) frequency.[6, 15]

Figure 2a shows peaks in the region between the G and G' bands (1650 – 2300 cm$^{-1}$), which are typically much lower in intensity compared to the other peaks in the Raman spectra of graphene. The strong dependence of peak frequencies and intensities of these modes on the number of layers can be observed clearly in Fig. 2a. The lowest frequency peak in Fig. 2a appearing at ~1750 cm$^{-1}$ is a double peak feature called the M band, which is an overtone of the out-of-plane o-TO phonon and has been observed in graphite and single-walled nanotube (SWNT) samples.[16, 17] The M band, which is activated by strong coupling between graphene layers, is suppressed for SLG and IBLG as observed previously.[14] In addition, the lower frequency peak in the M band (M$^-$) is downshifted by ~20 cm$^{-1}$ in BLG compared to FLG or HOPG (vertical dashed line in Fig. 2a). The peak at ~1860 cm$^{-1}$ in SLG has been assigned to a combination of the in-plane iTA phonon and the LO phonon and can be called the iTALO$^-$ mode.[14, 16] However, instead of a single peak as reported in previous studies,[14, 16] we observe a two-peak structure for this mode. Furthermore, the intensity of both peaks clearly decreases with increasing layers in graphene. We also observe these peaks in SLG samples on other substrates such as mica and quartz, confirming that the peaks are intrinsic to graphene and not a substrate effect. The third set of peaks in the range shown in Fig. 2(a) occur due to combinations between the iTO + LA (lower frequency peak) and LO + LA phonons (higher frequency peak).[14] It has recently been shown that the higher frequency LOLA peak is more sensitive to defects and decreases in intensity upon heat treatment.[18, 19] Finally a previously unidentified peak at ~2220 cm$^{-1}$ is observed in all graphene samples and its origin is discussed below.

Two novel features can be observed from Fig. 2a. A new mode appears at ~1880 cm$^{-1}$ (iTALO$^+$) as a shoulder on the higher frequency side of the iTALO$^-$ peak in SLG. In addition, this new mode is greatly suppressed in IBLG in contrast with SLG and BLG, indicating that it is very sensitive to the stacking order of graphene layers. As such, we refer to the absence of this mode as an indicator for IBLG. We tentatively assign this peak to a combination of the oTO and LO phonons around the K point of the graphene Billouin zone, as explained below. The second new feature in Fig. 2a is the appearance of a peak at ~2220 cm$^{-1}$ that has not been seen previously in graphene samples. This peak has, however, been observed in single-walled carbon nanotubes (SWNTs)[20] and is tentatively assigned as combination of the iTA and iTO phonons around the K point in the graphene Billouin zone.

Fig. 2b plots the frequencies of the iTALO$^-$, iTALO$^+$, iTOLA, LOLA, and iTOTA modes for SLG, IBLG, BLG, FLG, and HOPG. The iTALO$^-$ mode is absent for HOPG. The peaks involving the LO phonon, namely the iTALO$^-$, iTALO$^+$, and LOLA peaks increase in frequency due to increasing layers, while the iTOLA peak at ~1970 cm$^{-1}$ remains more or less at the same position. In addition, the frequency of the iTOTA mode at ~2220 cm$^{-1}$ (inset in Fig. 2b) is observed to decrease with increasing graphene layers. The frequency increases of the iTALO$^-$ and iTALO$^+$ modes indicate a high degree of sensitivity of these modes to the stacking order of graphene layers. All the combination modes in IBLG are further upshifted in frequency compared to both SLG and BLG. This is different from what is observed for AB-stacked graphene layers, suggesting an additional mechanism such as compressive strain between the two incommensurate graphene layers that causes an added stiffening of all the combination modes in IBLG compared to SLG and AB-stacked graphene layers. We confirmed that the relative shift of all the combination modes is maintained between the unfolded SLG and IBLG regions on the same sample, indicating that the results shown in Fig. 2b are not due to variations in electronic doping of different samples.

Interestingly, we find that all the combination mode frequencies exhibit an almost linear dependence on $1/n$ according to the following relation: $\omega(n) = \omega(\infty) + \beta/n$, where $n$ is the number of graphene layers, and $\beta$ is a constant (Fig.3). Such a linear dependence on $1/n$ has been observed previously for the G band phonons in exfoliated graphene.[10] As seen in Fig. 3, the values of $\beta$ for the iTALO$^-$ and LOLA are comparable to shifts caused by the van der Waals interactions (~12-13 cm$^{-1}$) in the radial breathing modes of bundled SWNTs,[21] indicating that these modes are mostly affected by layer stacking rather than changes in the electronic band structure. On the other hand, the high $\beta$ value for the iTALO$^+$ mode, which occurs due to a higher frequency shift with increasing graphene layers suggests that this mode maybe more sensitive to the electronic structure of graphene, similar to the G' band.

The dispersion of the combination modes discussed above versus laser energy is shown in Fig. 4. The iTOLA and LOLA modes upshift with laser energy by 204 cm$^{-1}$/eV and 223 cm$^{-1}$/eV respectively. These dispersions are similar to the peak dispersions of the iTOLA and LOLA modes in graphite and SWNTs.[17, 19, 22] In addition, the dispersion of the iTALO$^-$ mode is ~ 140 cm$^{-1}$/eV, similar to the value reported recently by Cong *et al.*,[14] while the dispersion of the iTALO$^+$ mode is a little higher (~150 cm$^{-1}$/eV). One could consider the two peaks around 1860 cm$^{-1}$ to occur in a similar fashion as the M band at ~1750 cm$^{-1}$, which also consists of two peaks. The two peak structure of the M band has been explained in the context of double resonance Raman scattering with the lower frequency (M⁻) peak attributed to scattering by a phonon with a momentum double that of the scattered electron ($q \approx 2k$), and the higher frequency (M⁺) peak due to scattering by a phonon with near zero momentum ($q \approx 0$).[17] This explains the fact that the M⁺ peak does not disperse with laser energy while the M⁻ peak downshifts with increasing laser energy. However, both the iTALO⁻ and iTALO⁺ modes are observed to shift with laser energy with similar dispersions, ruling out the $q \approx 0$ phonon within the framework of double resonance theory. Furthermore, the relative intensity between the iTALO⁻ and iTALO⁺ modes changes dramatically for IBLG. The inset in Fig. 4 shows the ratio of peak intensities of the iTALO⁺ and iTALO⁻ ($I_{iTALO+}/I_{iTALO-}$) modes plotted for SLG, IBLG, BLG and FLG. An obvious decrease in the ratio for IBLG can be observed, suggesting that the iTALO⁺ peak is quite sensitive to the interlayer interaction of individual graphene layers. A recent theoretical study predicted the absence of infrared modes in non AB-stacked graphene.[23] The suppression of the iTALO⁺ mode in IBLG could therefore occur due to the involvement of the infrared active oTO phonon. In fact, for the iTALO⁺ peak in HOPG at ~ 1940 cm⁻¹ (see Fig. 2b), a good agreement can be found for a combination of the oTO (~620 cm⁻¹) and LO modes (~1350 cm⁻¹) around the K point of the graphene Brillouin zone.[8] In addition, for the excitation ranges used in this study, the dispersions of the oTO and LO modes around the K point are both positive and could account for the ~150 cm⁻¹/eV dispersion of the iTALO⁺ mode. Based on the above arguments we tentatively assign the iTALO⁺ mode as a combination of the oTO and LO phonons around the K point of the graphene Brillouin zone. It is worth mentioning that second order modes with large dispersions (such as the iTOLA and LOLA modes) typically occur due to the combination of an acoustic and optical phonon, yet these other modes do not fit our data. Further theoretical and experimental studies are needed to understand why this particular combination mode appears for single layer and AB-stacked graphene but not IBLG.

The second previously unidentified mode in SLG at ~2220 cm$^{-1}$ (for $E_{laser}$ = 2.33 eV) has a negative dispersion with laser energy and the peak frequency downshifts with increasing excitation energy by ~ -56 cm$^{-1}$/eV. A peak at ~2200 cm$^{-1}$ has been observed in SWNTs but was left unassigned.[20] Moreover, as shown in Fig. 2b, the peak at ~2200 cm$^{-1}$ downshifts in frequency with increasing graphene layers in contrast to the other modes involving the LO phonon. The iTA branch around the K point has a negative dispersion and peaks around 1100 cm$^{-1}$ corresponding to the iTA phonon have been observed in graphite whiskers[16] and carbon nanotubes.[24] However, the dispersion of the iTA branch is ~ -75 – 100 cm$^{-1}$/eV,[25] which implies that the dispersion of its overtone would be twice as much. This makes it unlikely for the 2220 cm$^{-1}$ peak to be the overtone of the iTA phonon. On the other hand, a combination of the iTA phonon (at ~940 cm$^{-1}$) and iTO phonon (~1350 cm$^{-1}$) around the K point could account the 2200 cm$^{-1}$ mode. Moreover, the positive dispersion of the iTO phonon (~50 cm$^{-1}$/eV) and slight negative dispersion of the iTA phonon (~-20 cm$^{-1}$/eV) around the K point of the graphene Brillouin zone. However, there is limited experimental data available for iTA phonon branch around the K point of graphene (or graphite)[8] and this ambiguity could account for the the -50 cm$^{-1}$|/eV dispersion observed for the 2220 cm$^{-1}$ mode. We thus assign this mode as a combination of the iTA and iTO phonon (hence the name iTOTA) around the K point of the graphene Brillouin zone. All the combination modes discussed above are listed for SLG and HOPG in Table 1. Also included in Table 1 are the phonon modes involved in the double resonance Raman scattering process for these combination modes.

**Conclusions**

In summary, we have observed changes in various combination modes in the Raman spectra of graphene that depend on the number and stacking of layers. The overtone of the infrared active oTO phonon, also called the M band disappears for SLG and non AB-stacked bilayer samples, indicating that the M band is strongly dependent on stacking order of graphene layers. In addition, the lower frequency peak within the M band (M$^-$ peak) downshifts by ~20 cm$^{-1}$ for BLG compared to FLG and HOPG. A peak at ~ 1860 cm$^{-1}$ is attributed to iTA + LO phonons, and its intensity is observed to decrease with increasing graphene layers. Moreover, the iTALO band can be deconvoluted into two peaks, with similar dispersions versus laser energy. The higher frequency peak at ~1880 cm$^{-1}$ has a similar dispersion as the iTALO band and shows a strong dependence on stacking order of graphene layers. This peak is assigned to a combination of the oTO and LO phonons around the K point in the graphene Brillouin zone. A peak at ~2200 cm$^{-1}$ is observed for all graphene samples and is assigned to a combination of the iTA and iTO phonons around the K point. The peak frequencies of all the combination modes involving the LO phonon are observed to increase linearly with increasing graphene layers, indicating a strong coupling of the LO phonon between graphene layers.

**Methods**

The graphene samples having various layers were prepared using the standard mechanical exfoliation method from HOPG on 280 nm SiO$_2$/Si substrates (see Fig. S2 and S3 in the Supporting Information for optical microscope images of the samples).[26] The presence of SLG, IBLG, BLG and few layer graphene (FLG) areas were confirmed by atomic force microscopy (AFM) and micro-Raman spectroscopy. Raman spectra were acquired with a Renishaw InVia Raman microscope using $E_{laser}$ = 1.96, 2.33, and 2.41 eV. The incident laser beam was focused by a 50x objective and the laser power on the samples was kept to a minimum to avoid heating. All the Raman spectra were normalized with respect to the G band intensity and were baseline corrected prior to Lorentzian lineshape analysis.

**Acknowledgement**

RR gratefully acknowledges funding from AFOSR and the National Research Council associateship program. The material provided by RT, JK, and DT, and MI is based upon work supported by the National Science Foundation under Grant No. DMR-0955625. MI was supported by the summer faculty fellowship program from the American Society for Engineering Education for Summer 2010. RP and AMR greatly acknowledge the support from U.S. AFOSR (No. FA9550-09-1-0384).

*Supporting Information Available*: Raman spectra in the D band region from IBLG and SLG showing the I band at 1374 cm$^{-1}$ in IBLG. Optical microscope images (50 x magnification) of the SLG, BLG, FLG, and IBLG samples used in this study. This material is available free of charge *via* the Internet at http://pubs.acs.org.

**References**

1. Geim, A. K.; Novoselov, K. S. *Nat Mater* **2007,** 6, (3), 183-191.

2. Geim, A. K. *Science* **2009,** 324, (5934), 1530-1534.

3. Novoselov, K. S.; Geim, A. K.; Morozov, S. V.; Jiang, D.; Katsnelson, M. I.; Grigorieva, I. V.; Dubonos, S. V.; Firsov, A. A. *Nature* **2005,** 438, (7065), 197-200.

4. Zhang, Y.; Tang, T.-T.; Girit, C.; Hao, Z.; Martin, M. C.; Zettl, A.; Crommie, M. F.; Shen, Y. R.; Wang, F. *Nature* **2009,** 459, (7248), 820-823.

5. Latil, S.; Meunier, V.; Henrard, L. *Physical Review B* **2007,** 76, (20), 201402.

6. Ni, Z.; Wang, Y.; Yu, T.; You, Y.; Shen, Z. *Physical Review B* **2008,** 77, (23), 235403.

7. Ferrari, A. C.; Meyer, J. C.; Scardaci, V.; Casiraghi, C.; Lazzeri, M.; Mauri, F.; Piscanec, S.; Jiang, D.; Novoselov, K. S.; Roth, S.; Geim, A. K. *Phys. Rev. Lett.* **2006,** 97, 187401.

8. Saito, R.; Jorio, A.; Souza Filho, A.; Dresselhaus, G.; Dresselhaus, M.; Pimenta, M. *Phys. Rev. Lett.* **2001,** 88, (2), 027401.

9. Ferrari, A. C.; Meyer, J. C.; Scardaci, V.; Casiraghi, C.; Lazzeri, M.; Mauri, F.; Piscanec, S.; Jiang, D.; Novoselov, K. S.; Roth, S.; Geim, A. K. *Physical Review Letters* **2006,** 97, (18), 187401.

10. Gupta, A.; Chen, G.; Joshi, P.; Tadigadapa, S.; Eklund, P. *Nano Lett* **2006,** 6, (12), 2667-2673.

11. Cançado, L.; Reina, A.; Kong, J.; Dresselhaus, M. *Phys. Rev. B* **2008,** 77, (24), 245408.

12. Malard, L.; Pimenta, M.; Dresselhaus, G.; Dresselhaus, M. *Physics Reports* **2009,** 473, (5-6), 51-87.

13. Gupta, A. K.; Tang, Y.; Crespi, V. H.; Eklund, P. C. *arXiv:1005.3857v1* **2010**.

14. Cong, C.; Yu, T.; Saito, R. *arXiv:1010.3391v1* **2010**.

15. Poncharal, P.; Ayari, A.; Michel, T.; Sauvajol, J. L. *Physical Review B* **2008,** 78, (11), 113407.

16. Tan, P.; Dimovski, S.; Gogotsi, Y. *Philosophical Transactions of the Royal Society of London. Series A: Mathematical, Physical and Engineering Sciences* **2004,** 362, (1824), 2289.

17. Brar, V.; Samsonidze, G.; Dresselhaus, M.; Dresselhaus, G.; Saito, R.; Swan, A.; Ünlü, M.; Goldberg, B.; Souza Filho, A.; Jorio, A. *Phys. Rev. B* **2002,** 66, (15), 155418.

18. Ellis, A. *The Journal of chemical physics* **2006,** 125, 121103.

19. Rao, R.; Reppert, J.; Zhang, X. F.; Podila, R.; Rao, A. M.; Talapatra, S.; Maruyama, B. **2010**, arXiv:1010.4714v1

20. Tan, P.; Tang, Y.; Deng, Y. M.; Li, F.; Wei, Y. L.; Cheng, H. M. *Applied Physics Letters* **1999,** 75, 1524.

21. Rao, A. M.; Chen, J.; Richter, E.; Schlecht, U.; Eklund, P. C.; Haddon, R. C.; Venkateswaran, U. D.; Kwon, Y. K.; Tom; aacute; nek, D. *Physical Review Letters* **2001,** 86, (17), 3895.

22. Fantini, C.; Pimenta, M.; Strano, M. *The Journal of Physical Chemistry C* **2008,** 112, (34), 13150-13155.

23. Jiang, J.-W.; Tang, H.; Wang, B.-S. *Physical Review B* **2007,** 77, (23).

24. Dresselhaus, M.; Eklund, P. *Advances in Physics* **2000,** 49, (6), 705-814.

25. Saito, R.; Jorio, A.; Souza Filho, A. G.; Grueneis, A.; Pimenta, M. A.; Dresselhaus, G.; Dresselhaus, M. S. *Physica B: Physics of Condensed Matter* **2002,** 323, 100.

26. Novoselov, K. S.; Jiang, D.; Schedin, F.; Booth, T. J.; Khotkevich, V. V.; Morozov, S. V.; Geim, A. K. *Proc Natl Acad Sci U S A* **2005,** 102, (30), 10451-3.

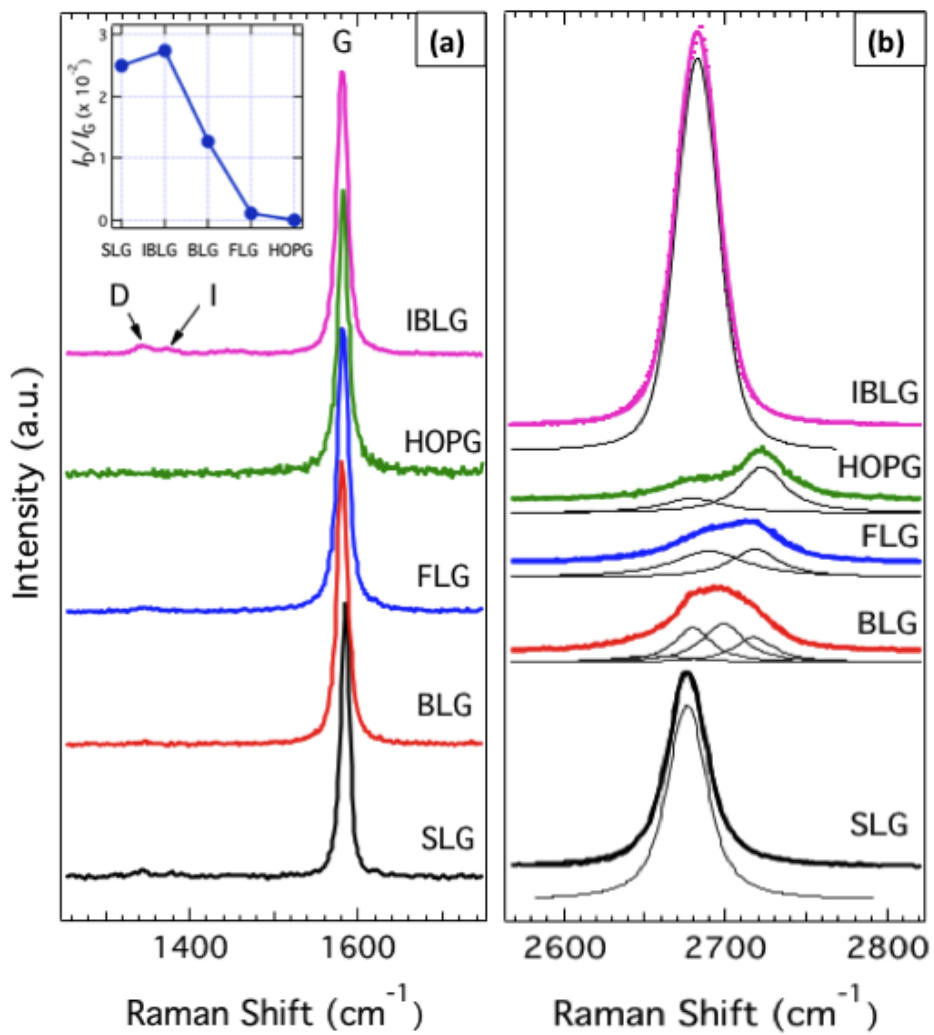

Figure 1. Raman spectra in the D and G band region (a), and the second order G' band region from single and multiple layer graphene samples (b). The inset in (a) shows the $I_D/I_G$ ratios for the various samples. All spectra are normalized with respect to the G band intensity and offset for clarity.

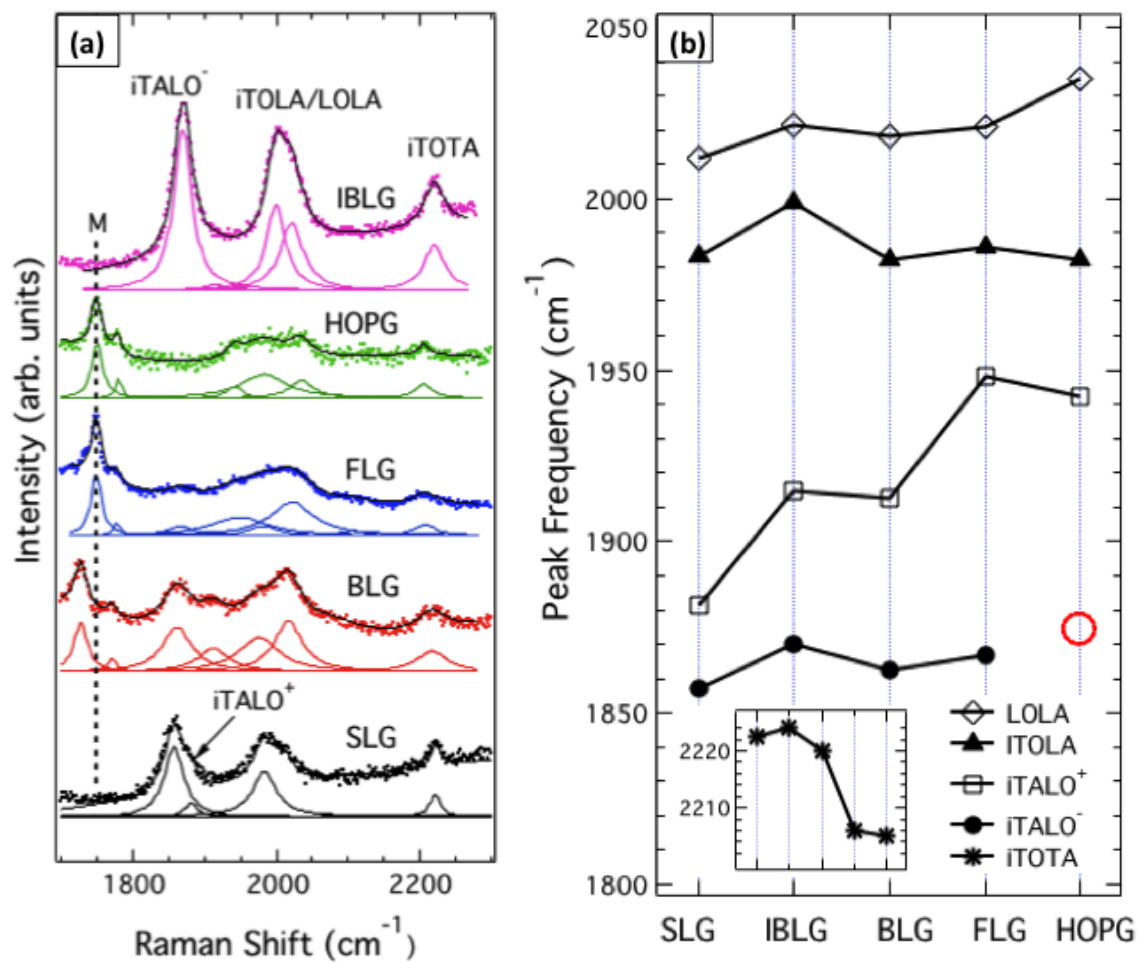

Figure 2. (a) Raman spectra between 1650 – 2150 cm$^{-1}$ from graphene samples collected with $E_{laser}$ = 2.33 eV. All spectra have been normalized by the G band intensity and fitted with Lorentzian peaks. (b) Change in peak frequency of the various second order double resonance Raman modes due to increasing layers in graphene samples. The absence of the iTALO peak at ~1860 cm$^{-1}$ in HOPG is indicated by the hollow (red) circle. Inset in (b): Position of the iTOTA peak (see text for discussion of peak assignment) for the various samples. The x-axis is same as in Fig. 2b.

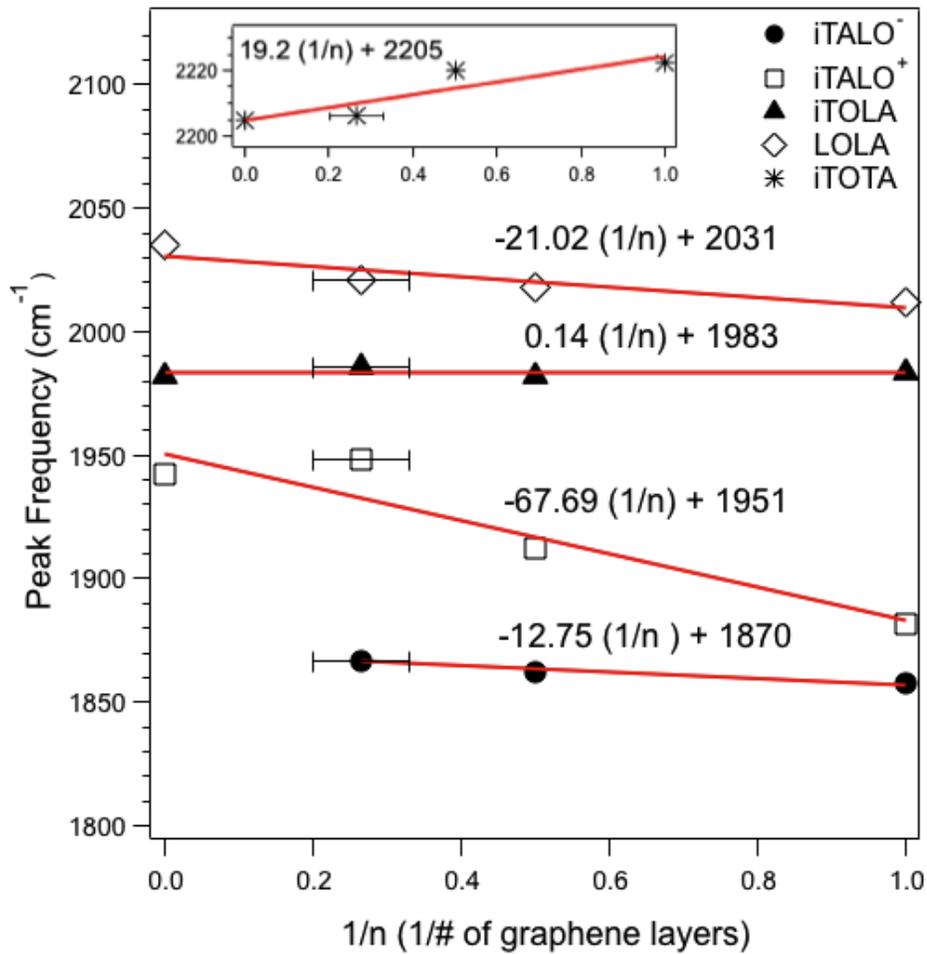

Figure 3. The peak frequencies of the iTALO-, iTALO+, iTALO, LOLA and iTOTA modes vs. $1/n$. The solid lines represent the results of a least squares fit to the data. The error bars for the FLG samples were obtained from AFM measurements which confirmed the presence of 3-5 graphene layers.

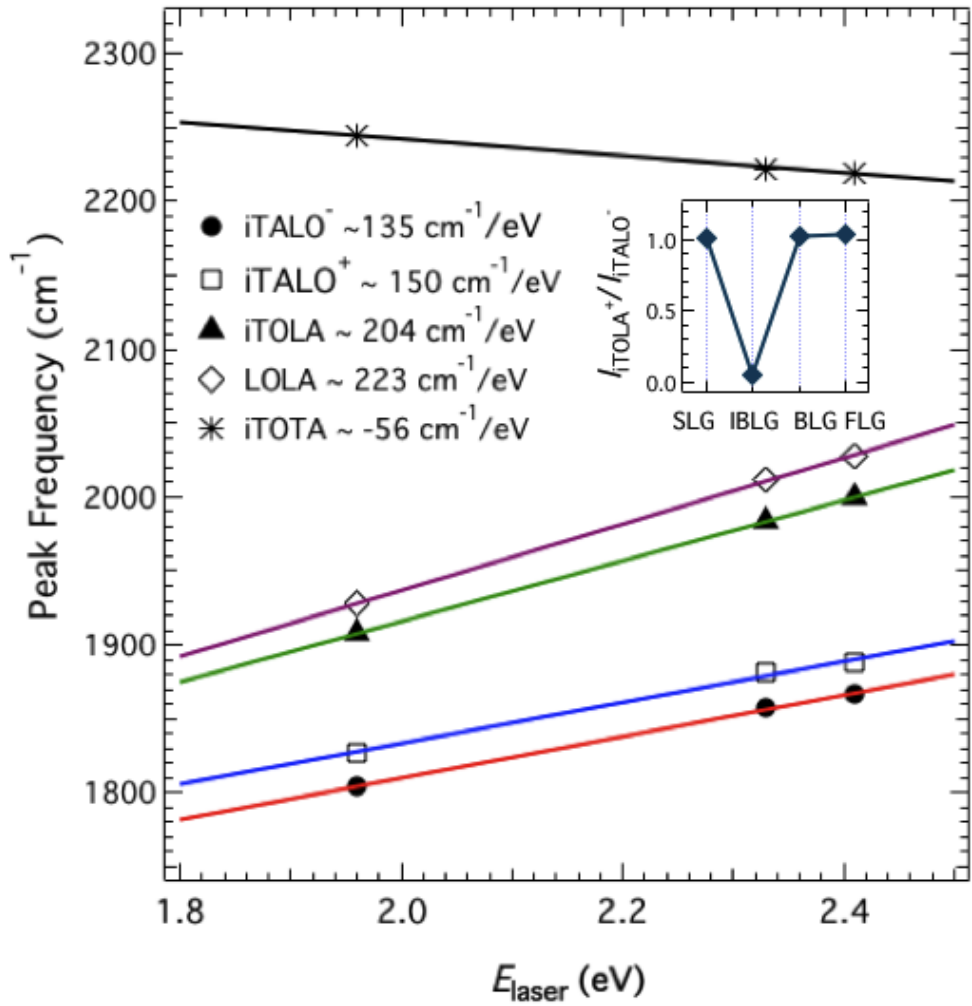

Fig. 4 Dispersion of the combination modes between 1650 – 2300 cm$^{-1}$ versus laser energy in SLG. Inset: Ratio of peak intensities of the oTOLO mode with respect to the iTALO mode for SLG, BLG, IBLG and FLG samples.

Table 1. Peak frequencies and assigned labels for experimentally observed double resonance Raman modes. The table also lists the peak dispersions versus laser energies for various combination modes for SLG and HOPG.

| Peak Frequencies (cm$^{-1}$) SLG (HOPG) | Mode | Phonons | Dispersion (cm$^{-1}$/eV) | Experimental Observation |
|---|---|---|---|---|
| (1725) | M | 2oTO | 0 | Refs. 16, 17 |
| (1750) | M | 2oTO | ~ -10 | Refs. 16, 17 |
| 1857 | iTALO$^-$ | iTA + LO | ~ 135 | This work and Ref. 14 |
| 1880 (1940) | iTALO$^+$ | oTO + LO | ~ 150 | This work |
| 1983 (1982) | iTOLA | LA + iTO | ~ 204 | Refs. 16, 18, 19, 22 |
| 2012 (2035) | LOLA | LA + LO | ~ 223 | Refs. 16, 18, 19, 22 |
| 2222 (2206) | iTOTA | iTA + iTO | ~ -56 | This work |